\documentclass[prd,twocolumn,a4paper,superscriptaddress]{revtex4}
\usepackage{graphicx,amssymb}
\begin{document}
\newcommand{\be}{\begin{equation}}
\newcommand{\ee}{\end{equation}}
\newcommand{\bq}{\begin{eqnarray}}
\newcommand{\eq}{\end{eqnarray}}
\newcommand{\bsq}{\begin{subequations}}
\newcommand{\esq}{\end{subequations}}
\newcommand{\bc}{\begin{center}}
\newcommand{\ec}{\end{center}}
\newcommand {\R}{{\mathcal R}}
\newcommand{\al}{\alpha}
\newcommand\lsim{\mathrel{\rlap{\lower4pt\hbox{\hskip1pt$\sim$}}
    \raise1pt\hbox{$<$}}}
\newcommand\gsim{\mathrel{\rlap{\lower4pt\hbox{\hskip1pt$\sim$}}
    \raise1pt\hbox{$>$}}}

\title{On the cosmological evolution of $\alpha$ and $\mu$ and the dynamics 
of dark energy}
\author{P. P. Avelino}
\email[Electronic address: ]{ppavelin@fc.up.pt}
\affiliation{Centro de F\'{\i}sica do Porto, Rua do Campo Alegre 687, 4169-007 Porto, 
Portugal}
\affiliation{Departamento de F\'{\i}sica da Faculdade de Ci\^encias
da Universidade do Porto, Rua do Campo Alegre 687, 4169-007 Porto, 
Portugal}

\date{22 April 2008}
\begin{abstract}

We study the cosmological evolution of the fine structure constant, $\alpha$, 
and the proton-to-electron mass ratio, $\mu=m_p/m_e$, in the context of a 
generic class of models where the gauge kinetic function is a linear function 
of a quintessence-type real scalar field, $\phi$, described by a Lagrangian with a 
standard kinetic term and a scalar field potential, $V(\phi)$. We further 
assume that the scalar field potential is a monotonic function of $\phi$ and that the 
scalar field is always rolling down the potential. We show that, for this 
class of models, low-redshift constrains on the evolution of $\alpha$ and 
$\mu$ can provide very stringent limits on the corresponding variations at 
high-redshift. We also demonstrate that these limits may be relaxed by 
considering more general models for the dynamics of $\alpha$ and $\mu$. 
However, in this case, the ability to reconstruct the evolution of the 
dark energy equation of state using varying couplings could be seriously 
compromised.

\end{abstract}
\maketitle

\section{\label{intr}Introduction}

Variations of $\alpha$ have been constrained over a broad redshift range 
($z = 0-10^{10}$) using various cosmological observations and laboratory 
experiments. The earliest constraints come from primordial nucleosynthesis 
which requires the value of $\alpha$ at $z \sim 10^{10}$ to be within a 
few percent of its present day value \cite{Bergstrom:1999wm,Avelino:2001nr,
Nollett:2002da} (although tighter 
constraints can be obtained for specific models 
\cite{Campbell:1994bf,Ichikawa:2002bt,Muller:2004gu,Coc:2006sx,Dent:2007et}). 
The cosmic microwave background temperature and polarization anisotropies 
give a constraint of comparable magnitude at much smaller redshifts 
$z \sim 10^3$ \cite{Avelino:2000ea,Avelino:2001nr,Martins:2003pe,Rocha:2003gc,
Stefanescu:2007aa}. 
 
At lower redshifts the situation is still controversial. A number 
of 
results, obtained through the measurement of the relative shifts of quasar 
spectral lines, suggest a cosmological variation of $\alpha$ and $\mu$ in 
the redshift range $z=1-4$ at about the $10^{-5}$ level \cite{Webb:1998cq,Murphy:2006vs,Ivanchik:2005ws,Reinhold:2006zn}. However other analysis have 
found no evidence for such variations \cite{Chand:2004ct,Molaro:2007kp,Wendt:2008fx}. This situation should be resolved in the next few years in particular 
with the next generation of high resolution spectrographs such as ESPRESSO 
planned for ESO's Very Large Telescope (VLT) which will be a stepping stone 
towards the CODEX spectrograph planned for the European Extremely large 
Telescope (E-ELT) \cite{Liske:2008zu}.

At even lower redshifts laboratory experiments at $z=0$ provide strong limits 
on $\alpha$ variability $|{\dot \alpha}/{\alpha}| = -2.6 \pm 3.9 \times 10^{-16} \  \rm yr^{-1}$ \cite{Peik:2006xy} while the constraints coming from the Oklo 
natural nuclear reactor limit the variation of $\alpha$ in the redshift range 
$z=0-0.2$ to be less than one part in $10^7$ \cite{Gould:2007au} assuming 
that only $\alpha$ has varied over time. Future laboratory tests will greatly 
improve current constraints. For example, the ACES (Atomic Clock Ensemble in 
Space) project will be able to constrain ${\dot \alpha}/\alpha$ at the 
$10^{-17} \  \rm yr^{-1}$ level \cite{ACES}. However, even more spectacular 
bounds (up to $10^{-23} \ \rm yr^{-1}$ \cite{Flambaum:2006yx}) may be available in 
the not too distant future. 

 On a more theoretical front, it was realized that in models where the 
quintessence field is non-minimally coupled to the electromagnetic field \cite{Carroll:1998zi,Olive:2001vz,Chiba:2001er,Wetterich:2002ic,Parkinson:2003kf,Anchordoqui:2003ij,Copeland:2003cv,Nunes:2003ff,Bento:2004jg,Avelino:2004hu,Doran:2004ek,Marra:2005yt,Avelino:2005pw} the dynamics of $\alpha$ 
is directly related to the evolution of the scalar field responsible for the 
dark energy. It was shown \cite{Avelino:2006gc}, that for a broad class of 
models, varying couplings may be used to probe the nature of dark energy over 
a larger redshift range than that spanned by standard methods (such as 
supernovae \cite{Perlmutter:1998np,Riess:1998cb,Riess:2004nr,Riess:2006fw} or weak lensing \cite{Huterer:2001yu}). 
Furthermore, it was claimed \cite{Avelino:2006gc} that a high-accuracy 
reconstruction of the equation of state may be possible all the way up to 
redshift $z \sim 4$.

Throughout this paper we shall neglect the spatial variations of $\alpha$ 
and $\mu$ which is usually a good approximation \cite{Shaw:2005gt,Avelino:2005pw}. These may be relevant in the 
context of chameleon-type models \cite{Brax:2004qh,Olive:2007aj} where masses and coupling constants are 
strongly dependent on the local mass density or if there are domain walls 
separating regions with different values of the couplings \cite{Menezes:2004tp}. However, in general, 
the late-time variation of the fundamental couplings is negligible 
in these models and consequently we shall not consider them 
further in this paper.

This paper is organized as follows. In Section II we shall 
consider a broad class of models for the evolution of $\alpha$ and $\mu$ 
where the gauge kinetic function is a linear function of a quintessence-type 
real scalar field described by a Lagrangian with a standard kinetic term and 
a scalar field potential, $V(\phi)$. 
We also assume that the scalar field 
potential is a monotonic function of $\phi$ and that the scalar field is 
always rolling down the potential. 
We show how low redshift observations 
can lead to very stringent constraints on the dynamics of $\alpha$ and $\mu$ at 
high redshifts, for models within this class. 
In section III we consider an even more general class of models where we 
relax one or more 
of the above assumptions and discuss the possible impact of this 
generalization on our ability to 
reconstruct the dark energy equation of state using varying couplings.
Finally we conclude in Section IV with a brief summary of our results and 
a discussion of future prospects. Throughout this paper we shall use fundamental 
units with $\hbar=c=G=1$ and a metric signature $(+,-,-,-)$.

\section{Dynamics of $\alpha$ and $\mu$}

In this section we shall consider a class of models described by the action
\begin{equation}\label{eq:L}
S=\int d^4x \, \sqrt{-g} \mathcal \, {\cal L} \, ,
\end{equation}
where $\mathcal L$ is the Lagrangian for a real scalar field $\phi$ coupled  
to the electromagnetic field with
\begin{equation} 
{\cal L} = {\cal L}_\phi + {\cal L}_{\phi F} + {\cal L}_{\rm other}\, , 
\end{equation} 
where
\begin{equation} 
{\cal L}_\phi= X - V(\phi)\, , 
\end{equation} 
\begin{equation}\label{eq:kinetic_scalar1}
X=\frac{1}{2}\nabla^\mu \phi \nabla_\mu \phi \,,
\end{equation}
\begin{equation} 
{\cal L}_{\phi F}= -\frac{1}{4}  B_F (\phi) F_{\mu \nu} F^{\mu \nu}\, , 
\end{equation}
$B_F(\phi)$ is the gauge kinetic function, $F_{\mu \nu}$ are the components of 
the electromagnetic field tensor and ${\cal L}_{\rm other}$ is the Lagrangian density of the other fields. The fine-structure constant is then given by 
\be 
\alpha(\phi)=\frac{\alpha_0}{B_F(\phi)} 
\label{gkfalpha} 
\ee 
and, at the present day, one has $B_F(\phi_0)=1$. 

The equation of motion for the field $\phi$ is
\begin{equation} 
\label{phieq} 
{\ddot \phi}+3H{\dot \phi}=-\frac{dV}{d \phi}-\frac{\alpha_0}{4 \alpha^2} 
\frac{d \alpha}{d \phi} 
F_{\mu \nu} F^{\mu \nu} \, , 
\end{equation} 
where a dot represents a derivative with respect to physical time, 
$H = {\dot a}/a$ and $a$ is the scale factor.
The time variations of the fine structure constant 
induced by the last term on the r.h.s. of Eqn. (\ref {phieq}) are very 
small (given Equivalence Principle constraints \cite{Olive:2001vz}) 
and can be neglected. Hence, throughout this paper we shall assume that the 
dynamics of $\phi$ is fully driven by the scalar field potential, $V(\phi)$ 
(and damped by the expansion). 

We will, for the moment, assume that the gauge kinetic function is a linear 
function of $\phi$ so that one has 
\be 
\frac{\Delta\alpha}{\alpha}=\beta \Delta \phi\,,
\label{gkfspec} 
\ee 
where $\Delta \alpha =\alpha_0-\alpha$, $\Delta \phi =\phi_0-\phi$, $\beta$ 
is a constant and we have also taken into account that 
$\Delta \alpha / \alpha \ll 1$ (at least for $z < 10^{10}$). We also 
assume that the scalar field 
potential, $V(\phi)$, is a monotonic function of $\phi$ and that the field 
$\phi$ is always rolling down the potential. If this is the case, 
and given a fixed value of ${\dot \phi}_0$, then $|\Delta \phi (z)| =
|\phi_0-\phi(z)|$ is maximized for a flat potential (here $z=1/a-1$ is the 
redshift). Note that if 
$dV/d\phi=0$ then the dynamics of the scalar field $\phi$ is simply given 
by
\begin{equation}  
{\ddot \phi}+3H{\dot \phi}=0
\end{equation}  
and, consequently, ${\dot \phi}={\dot \phi}_0 a^{-3}$.
For a non-flat monotonic potential $|\dot \phi|$ cannot increase so 
rapidly with redshift and so
\begin{equation}  
{\dot \phi}={\dot \phi}_0 a^{-3 s(a)}
\end{equation}  
with $s  \le  3$.
Note that, in this case, the contribution of the damping term 
due to the expansion of the universe is attenuated by the driving 
term due to the potential $V(\phi)$. Hence, given a fixed value of the 
kinetic energy of 
the scalar field $\phi$ at the present time its kinetic energy at 
$z > 0$ will always be smaller than the corresponding value in the flat 
potential case.
We may now calculate the 
value of $\Delta \phi (z)=\phi_0-\phi(z)$ for this special model (characterized by $dV/d\phi=0$) thus 
constraining the maximum allowed variations of $\alpha$ as a function of $z$.  
For $z < z_{\rm eq}$ one has $a \sim (t/t_0)^{2/3}$ and
\begin{eqnarray}  
f(z) &\equiv& \frac{\Delta \alpha (z)}{{\dot \alpha}_0 t_0} = \frac{\Delta \phi (z)}{{\dot \phi}_0 t_0}  = \frac{1}{{\dot \phi}_0 t_0}\int^{t_0}_t {\dot \phi} dt'\nonumber \\  
&=& \frac{3}{2} \int_a^1  u^{-5/2} du = (1+z)^{3/2}-1\,. 
\end{eqnarray}
If $z > z_{\rm eq}$ then $a \sim (t_{\rm eq}/t_0)^{2/3} (t/t_{\rm eq})^{1/2}$ 
and
\begin{eqnarray}
f(z) &=& 2 (1+z_{\rm eq})^{-1/2} \int_a^{a_{\rm eq}}  u^{-2} du + \frac{3}{2} \int_{a_{\rm eq}}^1  u^{-5/2} du \nonumber \\ 
 &=& 2 (1+z_{\rm eq})^{-1/2}(z-z_{\rm eq}) + (1+z_{\rm eq})^{3/2}-1 \,. 
\end{eqnarray}
Here, we have  assumed a sharp transition from the radiation to the 
matter-dominated era and we have neglected the small period of dark energy 
domination around the present time. This has a negligible impact 
on our results and greatly simplifies the calculations. 

We use the values $z_{\rm eq}=3200$ and $t_0=13.7 \, \rm Gyr$ consistent with 
latest WMAP 5-year results \cite{Komatsu:2008hk}. Note that at $f(z=4) \sim 10$, $f(z=10^3) \sim 3 
\times 10^4$ and $f(z=10^{10}) \sim 4 \times 10^8$. We thus see that a constraint on 
the value of ${\dot \alpha}_0 t_0/\alpha_0$ at $z \sim 0$ of about 1 part in 
$10^7$ consistent with no variation is enough to either rule out all current positive results for the 
variation of $\alpha$ or the broad class of varying $\alpha$ models 
presented above. On the other hand, low redshift 
constraints at the level of 1 part in $10^7$ or less will beat 
present CMB results at constraining the value of $\alpha$ around the 
recombination epoch (note that this level of precision will be within reach of 
the ACES project). Although even better constraints are needed in order to 
put useful bounds on the value of $\alpha$ at the nucleosynthesis epoch we 
should bear in mind that spectacular improvements may be expected in  
not too distant future \cite{Flambaum:2006yx,Cingoz:2006xf}. For example, 
Flambaum \cite{Flambaum:2006yx} has claimed that an improvement in the 
precision up to $10^{-23} \ \rm yr^{-1}$ (equivalent to a constraint of about 1 part 
in $10^{13}$ in ${\dot \alpha}_0 t_0/\alpha_0$) may be possible using the effect of the 
variation of $\alpha$ on the very narrow ultraviolet transition between the ground state and the first excited state in $^{299} \rm Th$ nucleus. If, in the future, we are 
able to achieve this level of precision and find a negative result for the variation of $\alpha$ then the values of $\alpha$ at recombination and nucleosynthesis would respectivelly have to be within about $10^{-8}$ and $10^{-4}$ of the present day value, a level of precision that cannot be easily achieved by other means.

The relation between the variations of $\alpha$ and $\mu$ is model dependent 
but, in general, we expect that
\begin{equation}
\frac{\dot \mu}{\mu}=R \frac{\dot \alpha}{\alpha}\,,
\end{equation}
where $R$ is a constant. The value of $R$ is of course model dependent (see \cite{Campbell:1994bf,Calmet:2001nu,Langacker:2001td,Calmet:2002ja,Olive:2002tz,Dine:2002ir,Dent:2008us} for a more detailed discussion of specific models) but if $|R|$ is large then variations of $\mu$ may well be easier to detect than variations of $\alpha$, a fact pointed out and studied in detail in ref. 
\cite{Avelino:2006gc}.

\section{More general models}

In the previous section we considered a class of models with 
${\cal L}(X,\phi)= X - V(\phi)$. In this section we consider an 
even more generic class of models with a real scalar field 
$\phi$ governed by an arbitrary Lagrangian of the form 
${\mathcal L}(X,\phi)$. Its energy-momentum tensor may be written 
in a perfect fluid form
\begin{equation}\label{eq:fluid}
T^{\mu\nu} = (\rho + p) u^\mu u^\nu - p g^{\mu\nu} \,,
\end{equation}
by means of the following identifications
\begin{equation}\label{eq:new_identifications}
u_\mu = \frac{\nabla_\mu \phi}{\sqrt{2X}} \,,  \quad \rho = 
2 X p_{,X} - p \, ,\quad p =  {\mathcal L}(X,\phi)\, .
\end{equation}
In Eq.~(\ref {eq:fluid}), $u^\mu$ is the 4-velocity field describing the motion of the fluid (for timelike $\nabla_\mu \phi$), while $\rho$ and $p$ are its proper energy density and pressure, respectively. The equation of motion for the scalar field is now
\begin{equation}
{\tilde g}^{\mu \nu} \nabla_\mu \nabla_\nu \phi=\frac{\partial{\mathcal L}}{\partial \phi}\,,
\end{equation}
where
\begin{equation}
{\tilde g}^{\mu \nu}=p_{,X} g^{\mu \nu} + p_{,XX}\nabla^\mu \phi \nabla^\nu \phi\,.
\end{equation}
An example of an algebraically simple but physically interesting class of 
Lagrangians is ${\mathcal L}(X) = f(X) - V(\phi)$ with $f(X) \propto X^n$. 
If the scalar field potential vanishes ($V=0$) then $w= 1/(2n-1)$ and 
consequently  when $n=1$ we have a standard massless 
scalar field, $n=2$ corresponds to background radiation and in the limit 
$n \rightarrow \infty$ the scalar field describes pressureless 
non-relativistic matter.
If the scalar field, $\phi$, is homogeneous then $X={\dot \phi}^2/2$ and its 
dynamics is given by
\begin{equation}
\label{eq:ddotphi}
n 2^{1-n} ({\dot \phi})^{2n-2}\left((2n-1){\ddot \phi} + 3 H{\dot \phi}\right)=-\frac{d V}{d\phi}\,,
\end{equation}
so that
\begin{equation}
{\dot \phi} = \phi_0 a^{-3/(2n-1)}\,,
\end{equation}
if $dV/d\phi = 0$. This is hardly surprising since for a constant $w$ the 
evolution of the energy density with the scale factor is given by 
$\rho =w X^n  \propto a^{-3(w+1)}=a^{-6n/(2n-1)}$. If $n > 1$ then the scalar 
field evolves more slowly with redshift than in the $n=1$ case and consequently the 
constraints considered in the previous section still apply here. However,
if $ 1/2 < n < 1$ then the scalar field may evolve much more rapidly than 
with $n=1$ and the above constraints may no longer be valid. We shall not 
consider models with $n < 1/2$ since, in this case, the sound speed squared, 
$c_s^2 = p_{,X}/\rho_{,X}$, is negative and consequently the solutions are 
unstable with respect to high frequency perturbations. 


We will now, for the sake of illustration, consider the evolution of $w$ 
for a family of models characterized by ${\mathcal L}(X) = f(X) - V(\phi)$ 
with $f(X) \propto X^n$ in two liming cases: case I - $|{\ddot \phi}| \ll 3 H 
|{\dot \phi}|/(2n-1)$ and case II - $|{\ddot \phi}| \gg 3 H |{\dot \phi}|/(2n-1)$.
Let us start with case I for which it is a good approximation to set 
${\ddot \phi}=0$ in Eq.~(\ref {eq:ddotphi}), so that 
${\dot \phi}={\rm constant}$. In this case 
\begin{equation}
\rho+p = 2n ({\dot \phi}^2/2)^n  = \rho_0+p_0 = 
(1+w_0)\rho_0\,,
\end{equation}
and consequently
\begin{equation}
w(a)=\frac{p}{\rho} = \frac{p_0 +\Delta V}{\rho_0 - \Delta V} = 
\frac{w_0 + \Delta V/\rho_0}{1 - \Delta V/\rho_0}\,,
\end{equation}
with $\Delta V = V_0-V$.
Also, we can show, using Eq.~(\ref {eq:ddotphi}) and the condition 
${\ddot \phi}=0$, that $\Delta V = - C \Delta \ln a = C \ln a$ where 
$C =  3 n 2^{1-n} {\dot \phi}^{2n} = 3 (1+w_0)\rho_0$ and we have taken $a_0=1$. The evolution of the equation of state is then given by
\begin{equation}
w(a)=\frac{w_0+3 (1+w_0)\ln a}{1-3 (1+w_0)\ln a}\,.
\end{equation}
Hence, we find no $n$ dependence in this limit. 

In case II with $|{\ddot \phi}| \gg 3 H |{\dot \phi}|/(2n-1)$ the energy 
density $\rho$ is approximately conserved and, to a good approximation, 
the equation of state parameter is simply given by
\begin{equation}
w(z)=-1 +\frac{2n}{\rho}\left(\frac{{\dot \phi}^2(z)}{2}\right)^n = -1 + (1+w_0) 
\left(\frac{\dot \phi (z)}{\dot \phi_0}\right)^{2n}\,.
\end{equation}
Consequently, in this limit, the evolution of $w$ with redshift, for a given 
evolution of $\phi$, is strongly dependent on $n$. Hence, if future 
constraints rule out 
the class of models described in the Section II then the ability to 
reconstruct the equation of state of dark energy from varying 
couplings would be compromised since to a 
given evolution of $\phi$ there may be many different possible evolutions 
for the equation of state (given fixed values for $\rho_0$ and $w_0$).

Of course, there are other possible generalizations to the class of models 
introduced in the previous section. For example, we could relax the 
assumption that the gauge kinetic function is a linear function of $\phi$. 
Then the dynamics of $\alpha$ would no longer need to be identical  
to that of $\phi$ and could even 
be very different from it. 
However, if we allow for an arbitrary gauge kinetic function we may no longer 
be able to use cosmological limits on the evolution of $\Delta \alpha/\alpha$ 
(or $\Delta \mu/\mu$) with redshift, $z$, to constrain the 
evolution of $\phi$. Consequently, the ability to reconstruct the 
equation of state of the dark energy would again be seriously compromised.

On the other hand, if the gauge kinetic function is linear in $\phi$ and 
${\cal L}(X,\phi)= X - V(\phi)$ we can, in principle, reconstruct the 
dark energy equation of state without further assumptions about $V(\phi)$. 
It may even be possible that future observations of the evolution of 
$\alpha$ (or $\mu$) with redshift require that 
$\dot \alpha$ (or $\dot \mu$) changes sign and lead us to consider non-monotonic potentials (see for example 
\cite{Fujii:2007zg}). However, such models will have to be 
fine-tuned in order to give an equation of state parameter $w \sim -1$ near 
the present time. Furthermore, it would be virtually impossible to determine 
whether the observed evolution of the couplings was due to special features of 
the scalar field potential or to a more complex kinetic term 
or gauge kinetic function. 
  
\section{\label{conc}Conclusions}

In this paper, in section II, we considered a generic class of models for 
the evolution of $\alpha$ and $\mu$. We then introduced a criterium that can 
be 
used to relate the limits on $\Delta \alpha/\alpha$ (or $\Delta \mu/\mu$) 
at different redshifts, for models within this class. 

We have demonstrated that low-redshift constraints on the 
evolution of $\alpha$ and $\mu$ can provide stringent limits on the 
corresponding variations at high-redshift. In particular, a constraint 
on the value of ${\dot \alpha}_0 t_0/\alpha_0$ at the $10^{-7}$ level 
(within reach of the ACES project) would, if consistent with no variation,  
be able to rule out all current positive results for the variation of $\alpha$.

We have also shown that future constraints at $z=0$ may lead to limits on 
$\Delta \alpha/\alpha$ at $z=1000$ which can be up to five orders of 
magnitude stronger than the best limits expected from future CMB experiments 
(such as Planck). At the nucleosynthesis epoch, $z=10^{10}$, the limits will 
be weaker by about $4$ orders of magnitude. Still, if an improvement up to 
$10^{-23} \ \rm yr^{-1}$ in the measurement precision of 
${\dot \alpha}_0 t_0/\alpha_0$ is obtained in the future then zero redshift 
constraints could lead to more stringent limits on the variation of 
$\alpha$ from $z \sim 10^{10}$ than the ones imposed by the observed light 
element abundances.

On the other hand, we have shown that if future observations lead us to 
adopt more general models, such as the ones studied in section III, then 
the above constraints can be relaxed. However, in this case we may no longer 
be able to trace the dynamics of dark energy using varying couplings.


\bibliography{alphaz}

\end{document}